\documentclass[aps,prl,twocolumn,groupedaddress,showpacs,showkeys]{revtex4}
\usepackage{graphicx}
\usepackage{array}
\usepackage{SIunits}
\usepackage{color} % %
\begin{document}

%\preprint{}

\title{Storage and recall of weak coherent optical pulses with an efficiency of 25\%}


\author{M. Sabooni, F. Beaudoin, A. Walther, Lin Nan, A. Amari, M. Huang, S. Kr\"{o}ll}
%\thanks{}
\affiliation{Department of Physics, Lund University, P.O.~Box 118, SE-22100 Lund, Sweden}


\date{\today}

\begin{abstract}
 We demonstrate experimentally a quantum memory scheme for the storage of weak coherent light pulses in an inhomogeneously broadened optical transition in a Pr$^{3+}$: YSO crystal at 2.1 K. Precise optical pumping using a frequency stable ($\approx$1kHz linewidth) laser is employed to create a highly controllable Atomic Frequency Comb (AFC) structure. We report single photon storage and retrieval efficiencies of 25\%, based on coherent photon echo type re-emission in the forward direction. The coherence property of the quantum memory is proved through interference between a super Gaussian pulse and the emitted echo. Backward retrieval of the photon echo emission has potential for increasing storage and recall efficiency.
\end{abstract}

%\keywords{}

\maketitle

\section{Introduction}
\par
Quantum information processing applications such as quantum networks require coherent and reversible mapping between light and matter \cite{Duan2001, Kimble2008}. Currently this requirement is expected to be met by means of quantum repeaters which allow temporal storage of quantum information in quantum memories and distribution of entanglement across long distances \cite{Briegel1998}. In the repeater protocol, the entanglement distribution has probabilistic behavior, and without quantum memories all probabilistic steps would have to succeed simultaneously. Quantum memories are therefore a key ingredient in any future long distance quantum communication scheme, and have thus been subject to very active development in recent time. To be useful in actual processing applications, these devices must maintain fidelities close to $100 \%$ during storage times of the order of seconds and also ensure high recall output efficiency \cite{Tittel2008,Riedmatten2008,Sangouard2008a}.
\par
Quantum memory protocols (such as Electromagnetically Induced Transparency (EIT) \cite{Harris1997} based on stopped light \cite{Longdell2005}, Controlled Reversible Inhomogeneous Broadening (CRIB) \cite{Moiseev2001,Nilsson2005a,Kraus2006} and Atomic Frequency Comb (AFC) \cite{Afzelius2009a,Riedmatten2008}, based on photon echo experiments) use atomic ensembles to obtain the required control and strong coupling between light field and the storage medium. These protocols can be adapted to the original DLCZ protocol \cite{Duan2001} to create entanglement generation via single photon detection. Therefore it is important to demonstrate that quantum memories can be operated at the single photon level.
\par
Protocols which enable storage of multiple temporal modes can increase the entanglement generation rate by a factor approximately equal to the number of temporal modes \cite{Simon2007}. Storing many modes with high efficiency requires a high optical depth, which generally can be an experimental difficulty. Achieving efficient multimode storage at lower optical depth is therefore desirable. It has been shown \cite{Afzelius2009a, Nunn2008}, that the number of modes N$_{m}$ that can be efficiently stored using EIT, scales as N$_m\approx\sqrt{\alpha L}$, where $\alpha L$ is the optical depth. CRIB offers a better scaling since N$_{m}\approx\alpha L$ (to double the number of modes it is sufficient to double $\alpha L$), but in the AFC protocol the number of temporal modes that can be stored with a given efficiency is independent of optical depth.
\par
High storage and retrieval efficiency in the single photon regime has been achieved previously using various protocols, e.g. in EIT by Eisaman et al \cite{Eisaman2005}, who published a storage and retrieval efficiency of 10 \% for $10 \mu{s}$ storage time in hot atomic vapor and also, by Choi et al \cite{choi2008}, who achieved 17 \% for 1$ \mu{s}$ in a cold atomic ensemble. Recently Lauritzen et al \cite{BjornLauritzen2009} published a storage and retrieval efficiency of 7\% for 600 ns in an erbium doped crystal using the CRIB protocol. Using the AFC protocol Riedmatten et al \cite{Riedmatten2008} obtained 0.5\% storage and retrieval efficiency after 250ns in Nd$^{3+}$:YVO$_{4}$ and Chaneli\'{e}re et al \cite{Chaneliere2009} achieved 9\% storage and retrieval efficiency in Tm$^{3+}$: YAG crystal.
\par
In this paper, we use the AFC protocol to store weak coherent pulses, with on the average 0.1 photons/pulse, up to 800 ns with 25\% storage and retrieval efficiency in a Pr$^{3+}$:YSO crystal. Generally three requirements must be fulfilled for an efficient quantum memory; 1) all light must be coherently absorbed, 2) the memory must maintain the coherence and amplitude information without dephasing and 3) the stored information must be reemitted. Compared to other ensemble based quantum memories, the AFC protocol is particularly strong on the first point. It achieves a high optical density over a large bandwidth, thus making it possible for the medium to efficiently absorb high bandwidth temporal pulse trains (multiple modes). Basically this is because the AFC memory in practice can be designed with a larger number of absorbers (atoms, ions) than can the other known ensemble based approaches. This larger number of atoms gives an increased light-matter interaction, potentially leading to higher efficiency in the AFC protocol. Recently, information was stored as a collective spin wave in the hyperfine ground state levels of Pr:YSO \cite{Afzelius2009b}, and coherence times up to 30 seconds have been achieved in these materials \cite{Fraval2005}.
\section{Materials \& Methodes}
\par

The experimental setup is shown in Fig.1. A 6-W Coherent Verdi-V6, 532 nm Nd:YVO$_4$ laser pumps a Coherent 699-21 dye laser which emits approximately 500 mW light at 605.82 nm. This frequency matches the $^3$H$_4$-$^1$D$_2$ transition of praseodymium ions doped into yttrium silicate (Pr$^{3+}$:Y$_2$SiO$_5$). 
\begin{figure}[ht]
    \includegraphics[width=8cm]{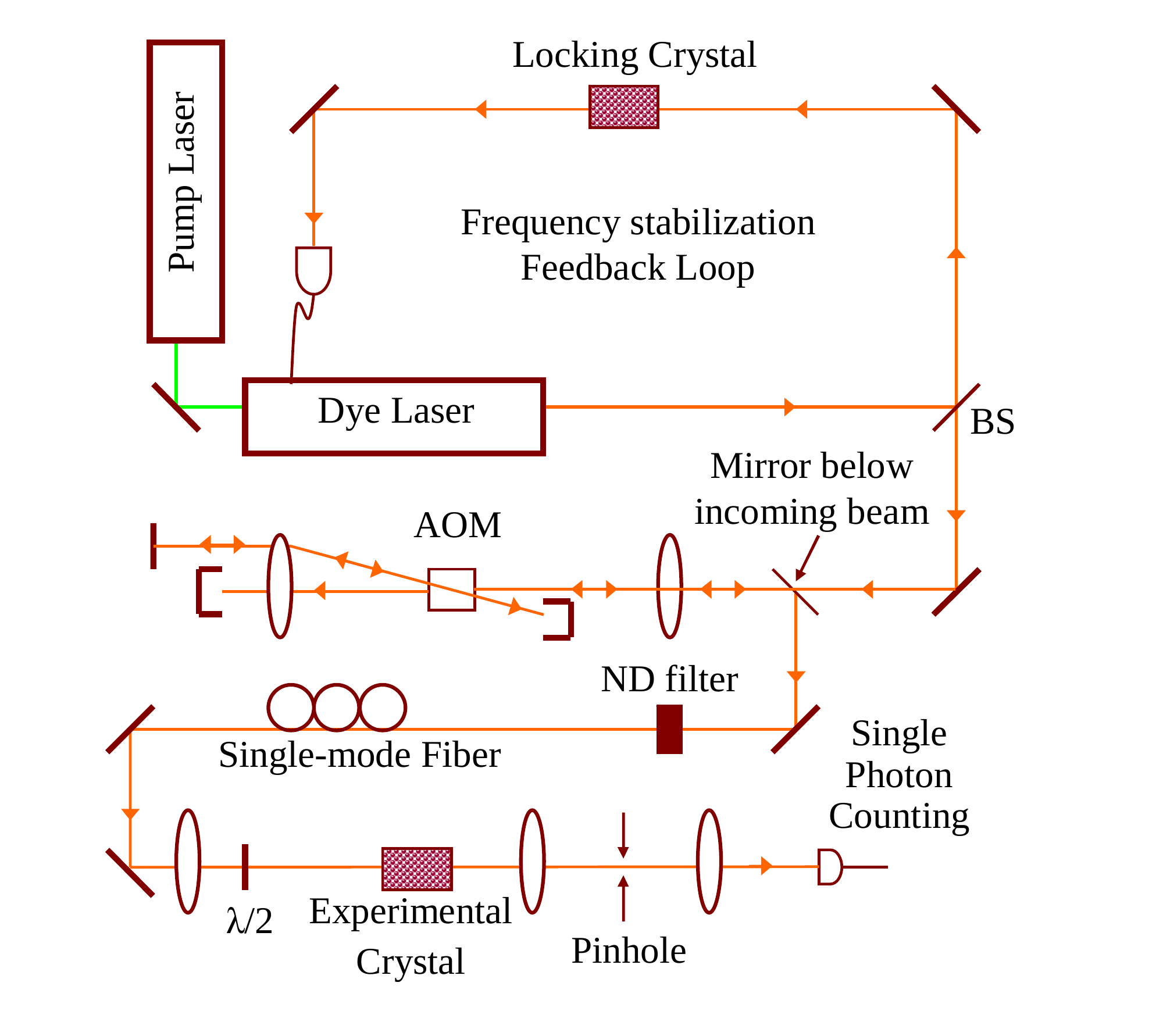}
    \caption{(Color online) Experimental set up. The frequency stabilization feedback system uses a Pr$^{3+}$:YSO crystal as locking crystal. All pulses are created using a double pass acousto-optic modulator (AOM). The incoming and outgoing beams in the AOM have different height and a mirror below the incoming beam redirects the outgoing beam to the experiment. Absorptive ND filters stands for most of the attenuation such that the AOM can operate in a linear light intensity vs RF amplitude regime for adjusting the storage pulse intensity. After spatial cleaning in the single mode fiber the light interacts with the Pr$^{3+}$:Y$_2$SiO$_5$ experimental crystal at 2.1 K. A $50 \mu m$ pinhole is used to select the center of the beam for the single photon counting detector.}
    \label{setup}
\end{figure}

We use a frequency stabilization feedback system \cite{Julsgaard2007}, based on a Pr$^{3+}$:Y$_2$SiO$_5$ crystal, which has a frequency drift below 1 kHz/s. An acousto-optic modulator (AOM) with 200 MHz center frequency, in a double pass arrangement that cancels spatial movement when changing the frequency of the diffracted beam, controls phase, amplitude, and frequency of the light. The double pass configuration also doubles the achievable frequency tuning range of the diffracted beam. The radio frequency signal used to drive the AOM is created by a 1-GS/s arbitrary waveform generator (Tektronix AWG520). Two mechanisms for light attenuation to the single photon level are used. The light beam can be attenuated by reducing the amplitude of the radio-frequency field driving the AOM, but since RF amplitude versus light amplitude is highly non-linear and difficult to calibrate due to low diffraction efficiency at low rf amplitudes, a shutter wheel with absorptive ND filters was added in the setup. The remaining control of the light amplitude was then handled through the AOM rf amplitude finally yielding about 0.1 photons per pulse at the sample. For cleaning up the spatial mode, after the AOM, the light was sent to a single mode fiber and then to a $\lambda$/2 plate such that the light polarization could be aligned along the transition dipole moment to yield maximum absorption. The sample is a Pr$^{3+}$:Y$_2$SiO$_5$ crystal with 0.05\% doping concentration and it has three perpendicular optical axes labeled D1, D2, and b, where the direction of light propagation was chosen along the b-axis. The crystal dimensions are 10$\times$10$\times$20mm along D1$\times$D2$\times$b axis. The width of the Gaussian beam at the center of the crystal is about $100\mu m$ and it is imaged (1:1) onto a $50\mu m$ pinhole. The pinhole gives us the opportunity to pick the light emitted by ions at the center of the beam that all have experienced roughly the same light intensity. Based on power measurement the maximum pinhole efficiency was 35\%. To keep the optical coherence time (T$_2$) of the praseodymium $> 100\mu s$, the crystal was immersed in liquid helium at 2.1 K. After the sample crystal a mechanical shutter was used to protect the detector from the intense AFC preparation pulses. A Hamamatsu single photon counting detector (Model H8259-01) with a quantum efficiency of 7.5\% and a minimum gating time of 100ns was used.
\begin{figure}[ht]
    \includegraphics[width=8.5cm]{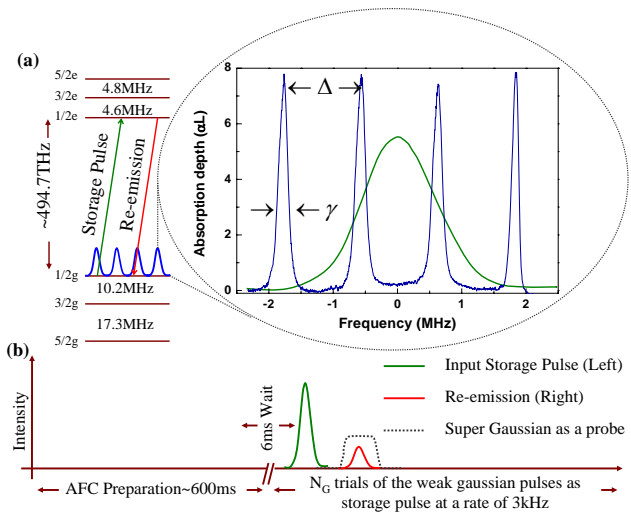}
    \caption{(Color online) a) The $^3$H$_4$-$^1$D$_2$ transition of the Pr$^{3+}$:Y$_2$SiO$_5$ crystal is used for the experiment. An experimental AFC absorption profile with defined height($\alpha$L), peak separation($\Delta$), and FWHM($\gamma$) with all ions in the state $\left|1/2g\right\rangle$ is created. A Gaussian pulse with center frequency set to the $\left|1/2g\right\rangle\rightarrow\left|1/2e\right\rangle$ transition is used as a storage pulse. b) A sequence of pulses which can be divided into preparation and storage pulses of a total duration of approximately 1.1 s is sent into the sample. A super Gaussian pulse is used as a probe pulse for interfering with the re-emission (echo) and showing the phase preservation of the AFC interface.}
    \label{AFC}
\end{figure}
\par
Let us now describe (see Fig. 2) in more detail how to create an Atomic Frequency Comb (AFC) \cite{Afzelius2009a}, i.e. a periodic series of narrow (FWHM=$\gamma$) and highly absorbing (optical depth $\alpha$L) peaks with a periodicity $\Delta$ as a frequency grating of ions in the $\left|1/2g\right\rangle$ state (see Fig. 2). By means of optical pumping, an 18 MHz zero absorption frequency region (see ref \cite{Atia}) is created within the 5 GHz wide Pr$^{3+}$:Y$_2$SiO$_5$ inhomogeneous profile \cite{Equall1995}. This non-absorbing frequency region is hereafter referred to as a spectral pit. Optimal pit creation needs a series of explicit pulses \cite{Atia}. Afterwards a frequency dependent absorption grating was created inside this spectral pit. Complex hyperbolic secant optical pulses were used to transfer some ions from $\left|5/2g\right\rangle$ to $\left|5/2e\right\rangle$ and then from the $\left|5/2e\right\rangle$ to $\left|1/2g\right\rangle$ state \cite{Rippe2005}. Changing the center frequency of the complex hyperbolic secant pulses, by an amount $\Delta$, is a simple technique to create a second peak inside the spectral pit. By repeating this procedure it is possible to create an atomic frequency comb in the $\left|1/2g\right\rangle$ level with a periodicity $\Delta$. The number of peaks (N) is limited by the $\left|1/2e\right\rangle\rightarrow\left|3/2e\right\rangle$ separation, ({$\delta \textit{f}_{\left|1/2e\right\rangle\rightarrow\left|3/2e\right\rangle}$), and $\Delta$ ($\Delta\times(N-1)<\delta \textit{f}_{\left|1/2e\right\rangle\rightarrow\left|3/2e\right\rangle}$).
\par
\section{Results \&Discussions}
\par
First a brief theoretical background to the AFC quantum memory interface. A weak light pulse is sent as a coherent state ($\left|\alpha\right\rangle_L$), on the average containing less than a single photon, with a well-defined bandwidth ($1/\tau_p$), to an AFC structure with well-defined peak separation ($\Delta$), peak width ($\gamma$), finesse $F=\Delta/\gamma$ and peak height ($\alpha$L) consisting of $N$ ions all in the $\left|1/2g\right\rangle$ state. The objective is to transfer all the energy of the weak coherent wave packet to the distribution of ions, $\left|g_1 \ldots g_N\right\rangle$,in the spectral grating with defined. After the light-matter interaction, the initial state of the N atoms will develop into the state \cite{Afzelius2009a}:
\begin{eqnarray}
\label{eq1}
\left|\alpha\right\rangle_A=\left|g_1\ldots g_N\right\rangle+\sum^N_{j=1}c_je^{i(\delta_jt-kz_j)}\left|g_1\ldots e_j\ldots g_N\right\rangle,
\end{eqnarray}
where $z_j$ is the spatial position of ion j, k is the wave number of the coherent single photon wave packet, $c_j$ is an amplitude that depends on the absorption probability, the spatial position and the absorption frequency of the ion. The AFC is a well-separated periodic structure with frequency detuning $\delta_j=m_j\Delta (m_j$ is an integer). Studying the time evolution of the superposition state in Eq. (\ref{eq1}) shows that in this state the ions will start to dephase. But the key feature of the AFC structure, with its periodicity $\Delta$, is that it gives rise to a constructive emission exactly after a time $t=1/\Delta$. The storage and retrieval efficiency of the AFC protocol can be written \cite{Afzelius2009a}:
\begin{eqnarray}
\label{eq2}
\eta =(1-e^{-\alpha L/F})^2e^{-7/F^2}
\end{eqnarray}
Due to the trade-off between the absorption (first) and dephasing (second) factors, there is an optimum value for the finesse as discussed in Ref. \cite{Afzelius2009a}.
\par
In our measurements we have two sets of pulses. They  correspond to the preparation pulses described in Ref. \cite{Atia} and the storage pulses described below. The amplitude of the storage pulse was adjusted taking into account attenuation factors due to windows, optical components, pinhole and the quantum efficiency of the single photon detector until the mean number of photons per pulse was about $\tilde{n}\approx 0.1$ at the sample. To improve measurement statistics, weak Gaussian storage pulses were sent in 2000 times at a rate of 3 kHz. To measure the efficiency, the storage pulse, which had a 200 ns FWHM, was sent through the empty spectral pit. In the empty pit there are no ions absorbing the pulse; therefore this is a good reference to compare with the recall pulse. The photon counting results fitted to a Gaussian time distribution for the emplty pit are shown (circles and dashed curve) in Fig. 3, respectively. The transmission is the same when the laser is tuned far outside the inhomogeneous absorption profile. Thus the pulse transmitted through the empty pit does not experience any absorption. 

\begin{figure}[ht]
    \includegraphics[width=8cm]{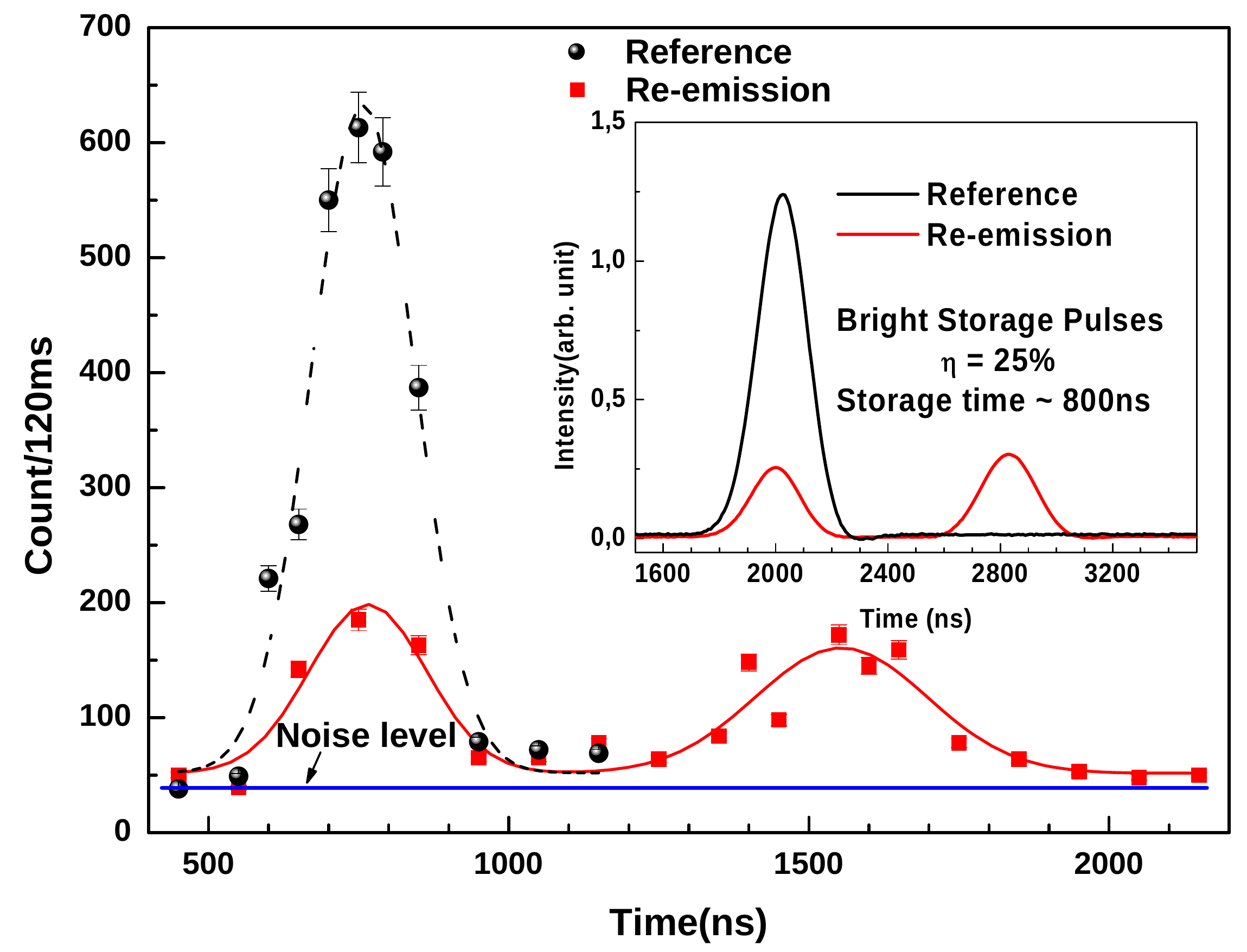}
    \caption{(Color online) Black circles (reference) are the single photon counts for the transmission of a 200 ns Gaussian storage pulse (n=0.1 photon per pulse) sent through the transparent empty pit without any AFC grating. Red squares correspond to the counts for an identical storage pulse sent through the AFC spectral grating with $\Delta =1.2$ MHz, $\gamma = 300$ kHz, and $\alpha L\approx$ 6. The ratio of the areas below the echo and the reference after fitting with a simple Gaussian function and subtracting the noise level is 0.25. The same result is achieved with bright storage pulses which is shown in the inset.}
    \label{echoAFC}
\end{figure}
\par
Next an AFC structure with peak separation $\Delta$=1.2MHz, peak width $\gamma$=200kHz, and $\alpha L\approx6$ is created and a Gaussian weak coherent storage pulse $(\tau_{FWHM}=200ns)$ tuned to the $\left|1/2g\right\rangle\rightarrow\left|1/2e\right\rangle$ transition (Fig 2a) is sent through the medium, which has a phase relaxation time much longer than the duration of the pulse. Re-emission (the echo) occurs after 800ns $(1/\Delta)$ (solid curve in Fig. 3). The measured storage and retrieval efficiency is 25\%, which means that the energy contained in the collective re-emission (echo) pulse relative to the energy transmitted through empty pit (dashed curve) is 0.25.
\par
So far we have considered the efficiency properties of our interface but for quantum memory applications it is vital that the interface also conserve the phase of the storage pulse with high fidelity. To show the phase preservation a super Gaussian (n=7) pulse with controllable phase, overlapping the re-emission pulse temporally, but with a 2.3 MHz frequency offset is prepared. This frequency offset allows us to send the super Gaussian pulse through a transparent frequency window inside the pit between $\left|1/2g\right\rangle\rightarrow\left|1/2e\right\rangle$ and $\left|1/2g\right\rangle\rightarrow\left|3/2e\right\rangle$ transitions peaks \cite{Atia}. To have enough frequency space between these two transitions, the peak separation $(\Delta)$ is reduced to 1MHz but the rest of the AFC parameters are the same as before. In these measurements the $\tau_{FWHM}$ of the Gaussian pulse is 420ns, whereas the super Gaussian full width at half maximum is 840ns.
\begin{figure}[ht]
    \includegraphics[width=8cm]{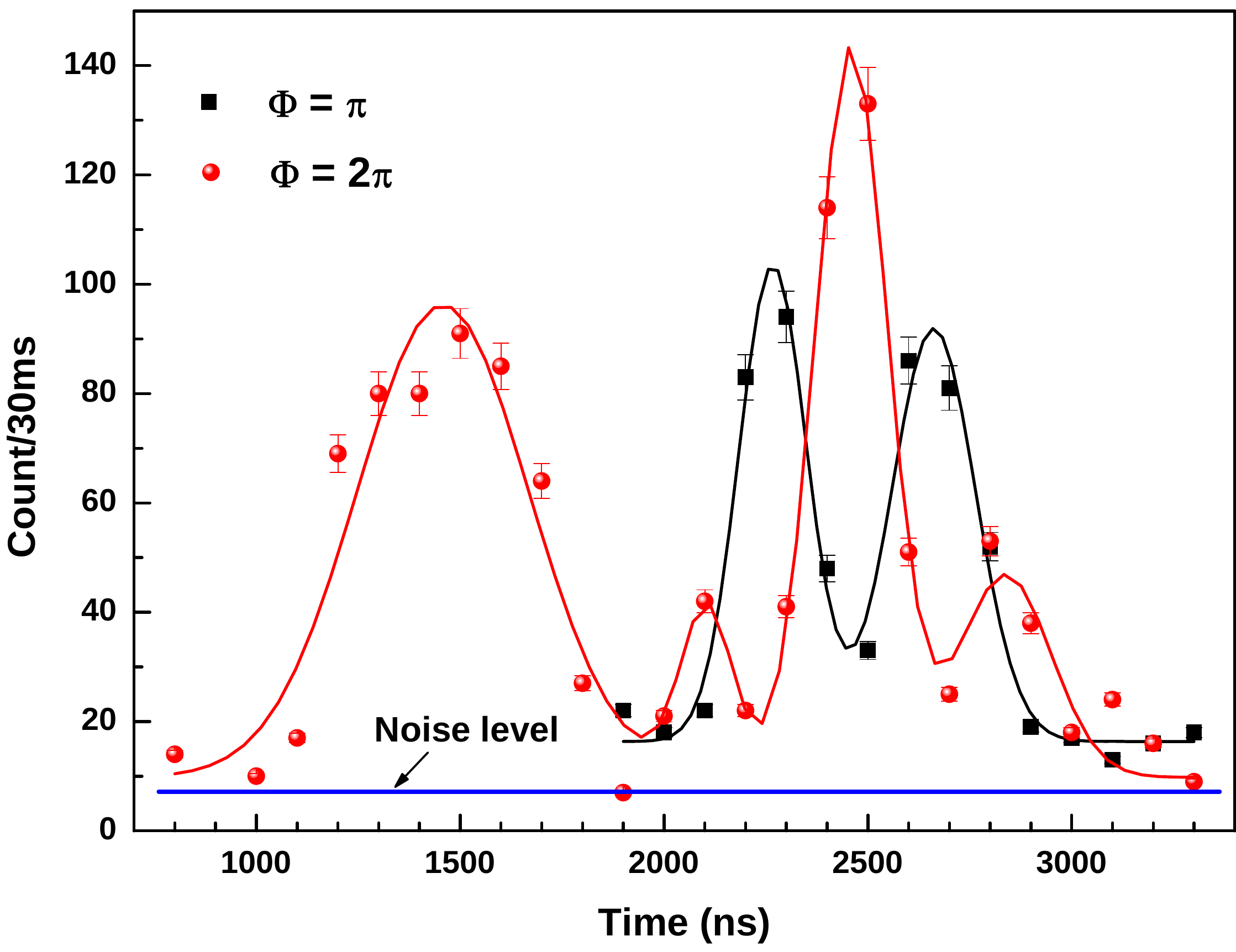}
    \caption{(Color online) Beating between re-emission (echo) and the super Gaussian (n=7) pulses. The first pulse to the left is the transmitted part of the input storage pulse, which is independent of the phase of the supergaussian pulse. The visibility after subtracting the noise level is 83\%. The interference patterns is inverted when the phase factor of super Gaussian pulse is changed by $\pi$, demonstrating phase conservation of the weak coherent storage pulse by the AFC spectral grating.}
    \label{coherencAFC}
\end{figure}
\par
Due to the interference between the two pulses, a beating pattern is expected at the detector at the time of the re-emission (the echo) pulse (Fig. 4). The measured visibility after subtraction of the noise level is 83\%. To show that the phase is conserved, the phase of the super Gaussian pulse was changed by $\pi$ and the beating pattern is reversed (Fig. 4).
\par
The 25\% storage and retrieval efficiency obtained in this paper is independent of the storage pulse intensity. Further, all results with the bright storage pulses \cite{Atia} are reproducible in the single photon regime. The experiment is based on coherent photon-echo type re-emission experiments in the forward direction. As discussed in Ref. \cite{Riedmatten2008} the maximum achievable efficiency for emission in the forward direction is 54\%. This is basically limited by the re-absorption of the emitted echo. To improve the efficiency beyond 54\% counter propagating control fields \cite{Moiseev2001}, or external control of the ion frequency \cite{Kalachev2007} could be used, alternatively, the crystal could be placed inside a cavity \cite{Afzelius_lund}. The efficiency could also be enhanced by using optimum input pulses \cite{Irina2007, Phillips2008, Kalachev2007}.
\begin{acknowledgments}
\par
This work was supported by the Swedish Research Council, the Knut and Alice Wallenberg Foundation and the European Commission through the integrated project QAP.
\end{acknowledgments}

\end{document}